\documentclass[showpacs,aps,graphicx,twocolumn]{revtex4}
\usepackage{amsmath}
\usepackage{amscd}
\usepackage{graphicx}
\begin{document}


\title{Deterministic polarization entanglement purification using spatial entanglement}
\author{ Xi-Han Li\footnote{
Email address: xihanli@cqu.edu.cn}}
\address{Department of Physics, Chongqing University,
Chongqing 400044,  China}

\date{\today }

\begin{abstract}
We present an efficient entanglement purification protocol with hyperentanglement, in which
additional spatial entanglement is utilized to purify the two-particle polarization entangled state.
The bit-flip error and phase-flip error can be corrected and eliminated in one step. Two remote parties can obtain maximally
entangled polarization states deterministically and only passive linear optics are employed.
We also discuss the protocol with practical quantum source and noisy channel.

\end{abstract}

\maketitle

\section{introduction}
The photon is destined to have a central role in quantum communication owing to its high-speed transmission and outstanding low-noise properties.
However, its polarization degree of freedom (DOF) is vulnerable to the noise in the quantum channel. For instance, the thermal fluctuation, vibration, the imperfection of an optical fiber and the birefringence effects will inevitably affect the polarization of photons.  To implement faithful quantum communication between
remote parties over noisy channel, many method have been proposed, one of which is entanglement purification protocol (EPP) \cite{1}.
The entanglement between photons is incident to attenuate during the transmission and then the quantum state becomes a less-entangled and mixed one.
EPP is a method to extract maximally entangled states by consuming less-entangled pairs using only local operations and classical communication.
The EPP, along with entanglement swapping \cite{swap} is also important element in quantum repeater for long-distance quantum communication \cite{repe}.
The first EPP was proposed by Bennett in 1996 \cite{1} and several improved schemes were present subsequently \cite{2,3,4,5,6,7}. In 2002, Simon et al present an
EPP works for currently available parametric down-conversion (PDC) sources \cite{5}, in which the spatial entanglement is first used as an additional resource
to purify polarization entanglement.
%
Generally speaking, the conventional EPPs can increase the entanglement and fidelity of quantum states step by step. But the quantum resources consumed also increase with the step exponentially.  And there is a minimum requirement on the fidelity of initial state to be purified.
More recently, the original deterministic entanglement purification protocol (DEPP) with hyperentangled state
 was proposed by Sheng et al \cite{17}, in which two additional DOFs are used to correct the bit-flip error and phase-flip error of the polarization entangled state in two steps,
 respectively. Two remote parties can get maximally entangled photon pairs in a deterministic way.
However, their protocol employed the cross-Kerr nonlinearity, which is difficult to realize at the single-photon level with current technique \cite{18}.

In this letter, we present a DEPP with less quantum resource. Only one addition DOF is required to purify the polarization entangled state and the bit-flip and phase-flip errors can be corrected and rejected in one step. Our set-up, which is composed of passive linear optics, is easily accessible with present technology. We also discuss this scheme with practical quantum source and transmission channel.

\section{ideal deterministic entanglement purification}
In the first place, we introduce the basic principle of our DEPP with an idea quantum source and a simple noise model, in which we only consider the impact of noise on the polarization state.  The hyperentangled state in polarization and spatial mode can be written as
\begin{eqnarray}
\vert \Psi \rangle_{AB}=\vert \Psi\rangle_p \vert \Psi \rangle_s=\frac{1}{2}(\vert HH \rangle +\vert VV \rangle)(\vert a_1b_1 \rangle +\vert a_2b_2 \rangle), \label{state}
\end{eqnarray}
which can be produced by Simon's design \cite{5}. Here $\vert \Psi \rangle_p$ and $\vert \Psi \rangle_s$ denote the polarization state and spatial one, respectively. And $H,V$ represent the horizontal and vertical photon polarization  and $a_1,b_1,a_2,b_2$ represent four spatial modes of photons.

As the spatial DOF is more stable than polarization, the entanglement in spatial modes is assumed to be invariant during the transmission. But the polarization state changes into a mixed one after transmission through noisy channels:
\begin{eqnarray}
\rho_p &=& \alpha \vert \Phi^+ \rangle_{AB} \langle \Phi^+ \vert + \beta \vert \Phi^- \rangle_{AB} \langle \Phi^- \vert \nonumber\\&&+\delta \vert \Psi^+ \rangle_{AB} \langle \Psi^+ \vert +\eta \vert \Psi^- \rangle_{AB} \langle \Psi^- \vert.
\end{eqnarray}
Here $\vert \Phi^\pm \rangle_{AB}=\frac{1}{\sqrt{2}}(\vert HH \rangle \pm \vert VV\rangle)_{AB}$ and $\vert \Psi^\pm \rangle_{AB}=\frac{1}{\sqrt{2}}(\vert HV \rangle \pm \vert VH\rangle)_{AB}$ are four Bell states.
These four parameters represent the proportion of each Bell states in the mixed state and $\alpha+\beta+\delta+\eta=1$.
The density matrix of the whole quantum state including two DOFs becomes $\rho =\rho_p \rho_s$, where
$\rho_s =\frac{1}{2}(\vert a_1b_1 \rangle +\vert a_2b_2 \rangle)(\langle a_1b_1 \vert +\langle a_2b_2 \vert)$.

The quantum system equals to a probabilistic mixture of four pure  states: the photon pairs in state $\vert \Phi^+ \rangle_f \vert \Psi \rangle_s$ with a probability $\alpha$, in state $\vert \Phi^- \rangle_f \vert \Psi \rangle_s$ with a probability $\beta$, and in $\vert \Psi^+ \rangle_f \vert \Psi \rangle_s$ ($\vert \Psi^- \rangle_f \vert \Psi \rangle_s$) with a probability $\delta$ ($\eta$).
 These three unwanted polarization states $\vert \Psi^+ \rangle_{AB}$, $\vert \Phi^- \rangle_{AB}$, $\vert \Psi^- \rangle_{AB}$ can be viewed as a bit-flip error, a phase-flip error and both bit-flip and phase-flip errors taking place, respectively. The aim of EPP is to correct these errors and make the two parties sharing the pure maximally entangled polarization state $\vert \Phi^+ \rangle_{AB}$.

\begin{figure}[!h]
\centering
\includegraphics[height=1.05in]{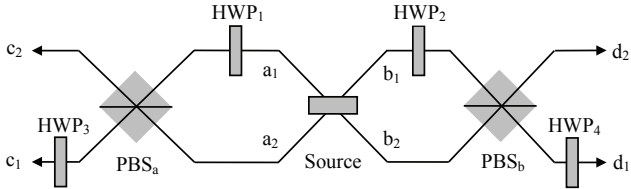}
\caption{The schematic diagram of the principle of deterministic entanglement purification.
The source produces the hyperentangled state and sends two photons to two remote parties. The two parties get the desired pure maximally entangled polarization state in four different spatial modes.}
\end{figure}

The principle of our DEPP is shown in Fig.1. The polarizing beam splitter (PBS) transmits the horizontal state and reflects the vertical one. And HWPs effect bit-flip operations as $\vert H \rangle \rightleftharpoons \vert V \rangle$. We discuss the faultless case first. The state $\vert \Phi^+ \rangle_f \vert \Psi \rangle_s=\frac{1}{2}(\vert HH \rangle +\vert VV \rangle)(\vert a_1b_1 \rangle +\vert a_2b_2 \rangle)$ evolves as
{\footnotesize\begin{eqnarray}
&\rightarrow&\frac{1}{2}(\vert HH \rangle_{a_1b_1}+\vert HH \rangle_{a_2b_2}+\vert VV \rangle_{a_1b_1}+\vert VV \rangle_{a_2b_2})\nonumber\\
&\xrightarrow[{\tiny HWP_2}]{{\tiny HWP_1}}&
\frac{1}{2}(\vert VV \rangle_{a_1b_1}+\vert HH \rangle_{a_2b_2}+\vert HH \rangle_{a_1b_1}+\vert VV \rangle_{a_2b_2})\nonumber\\
&\xrightarrow[{\tiny PBS_b}]{{\tiny PBS_a}}&
\frac{1}{2}(\vert VV \rangle_{c_2d_2}+\vert HH \rangle_{c_2d_2}+\vert HH \rangle_{c_1d_1}+\vert VV \rangle_{c_1d_1})\nonumber\\
&\xrightarrow[{\tiny HWP_4}]{{\tiny HWP_3}}&
\frac{1}{2}(\vert HH \rangle +\vert VV \rangle)(\vert c_1d_1 \rangle +\vert c_2d_2 \rangle).
\end{eqnarray}}
We find the final polarization state is identical with the
initial one and the two parties can get the maximally entangled polarization state $\vert \Phi^+ \rangle$ in modes $c_1d_1$ or $c_2d_2$, which means the photon $A$ emits from $c_1$($c_2$) while photon $B$ in $d_1$($d_2$) accordingly.

For the case of a phase-flip error, the state $\vert \Phi^- \rangle_f \vert \Psi \rangle_s=\frac{1}{2}(\vert HH \rangle -\vert VV \rangle)(\vert a_1b_1 \rangle +\vert a_2b_2 \rangle)$ becomes $\frac{1}{2}(\vert HH \rangle +\vert VV \rangle)(\vert c_2d_2 \rangle -\vert c_1d_1 \rangle$.
The phase-flip error is transferred from the polarization DOF to the spatial one, which has no effect on the final state. The polarization state obtained by the two parties is the same as the faultless case. In other words, two parties will get the maximally entangled state $\vert \Phi^+ \rangle$ in modes $c_1d_1$ or $c_2d_2$ too.

Considering another two error situations, the states $\vert \Psi^\pm \rangle_f \vert \Psi \rangle_s=\frac{1}{2}(\vert HV \rangle \pm \vert VH \rangle)(\vert a_1b_1 \rangle +\vert a_2b_2 \rangle)$ change into $\frac{1}{2}(\vert HH\rangle +\vert VV \rangle)(\vert c_2d_1 \rangle \pm \vert c_1d_2 \rangle$.
Both the bit-flip and phase-flip errors are transferred to the spatial DOF, and the bit-flip error is corrected by HWP$_3$ and HWP$_4$. The two parties can get $\vert \Phi^+ \rangle$ in modes $c_2d_1$ or $c_1d_2$.

Take these four situations discussed above into account, we find the essence of DEPP is transferring errors from the polarization DOF to the spatial one. The bit-flip error is corrected with the help of spatial modes, and the phase-flip error will be eliminated in the process of the spatial measurements. During the purification, the original quantum system composed of a mixed polarization state and a pure spatial state was changed into a product state of a pure polarization state $\vert \Phi^+ \rangle$ and a mixed spatial modes. Thus the spatial state can be seen as a superposition of four different spatial modes $c_1d_1, c_2d_2, c_1d_2$ and $c_2d_1$, into which the state collapses during the measurement. The purified pure maximally entangled polarization state $\vert \Phi^+ \rangle$ is accessible by postselection. The entanglement is useful in quantum communication although in a postselected sense. For example, four detectors set in the four output ports can be used to implement the BBM92 quantum key distribution protocol \cite{bbm92}. This protocol is deterministic as no case is discarded. In a word, the error free maximally entangled polarization state $\vert \Phi^+ \rangle$ can always be shared between two remote parties with one step purification.

\section{Deterministic Entanglement purification in practical situations}

In Sec. II, we described the basic principle of our DEPP in an ideal case, in which the spatial DOF do not suffer from the noise.
In fact, the spatial entanglement would also be affected by the environment, but the effect is less than polarization\cite{effect}. Generally, the probability for bit-flip errors in spatial modes is extremely low, but the phase would be
influenced by the phase instability of environment and length difference between different spatial modes.
In 2008, experiment showed the phase in long fibers  remains stable on the order of 100 $\mu$s in a realistic situation\cite{phase}. Therefore, the relative phase is mainly caused by path-length dispersion. After the transmission, not only the polarization states change, but the spatial state may also become $\frac{1}{\sqrt{2}}(\vert a_1b_1\rangle + e^{i\phi}\vert a_2b_2 \rangle)$, where $\phi=k\Delta L$. Here $k$ is the wave vector of photons and $\Delta L$ is the path-length difference between two spatial modes \cite{17}. Then after the purification, the polarization entangled state may becomes $\frac{1}{\sqrt{2}}(\vert HH \rangle + e^{i\phi}\vert VV \rangle)$ for spatial modes $c_1d_1$ and $c_1d_2$ and $\frac{1}{\sqrt{2}}( e^{i\phi}\vert HH \rangle +\vert VV \rangle)$ for modes $c_2d_2$ and $c_2d_1$. As the path length is fixed, the two parties can get pure entangled states via different spatial modes, and obtain the standard Bell state $\frac{1}{\sqrt{2}}(\vert HH \rangle + \vert VV \rangle)$ with phase compensations.

In the preceding discussion, we supposed the quantum source was an ideal one, which creates exactly one hyperentangled pair each time. However,  due to the quasithermal character of down-conversion, there exists the probabilities for one source to emit two pairs considering the practical source. In Ref.\cite{5}, a pump light traverses a crystal twice, whose situation can be described approximately by the Hamiltonian
\begin{eqnarray}
H&=&\gamma [(a^\dag_{1H}b^\dag_{1H}+a^\dag_{1V}b^\dag_{1V})+re^{i\varphi}(a^\dag_{2H}b^\dag_{2H}+a^\dag_{2V}b^\dag_{2V})]\nonumber\\
&&+\texttt{H.c.},
\end{eqnarray}
where $r$ denotes the relative probability of emission of photons into the lower modes $a_2b_2$ compared to the upper modes $a_1b_1$ and $\varphi$ is the phase between these two possibilities. We assume ideal conditions to set $r=1$ and $\varphi=0$. When the source prepares a single-pair as $(a^\dag_{1H}b^\dag_{1H}+a^\dag_{1V}b^\dag_{1V}+a^\dag_{2H}b^\dag_{2H}+a^\dag_{2V}b^\dag_{2V})\vert 0\rangle$, the state is equal to Eq. (\ref{state}) in a different notation, which can be purified perfectly by our protocol.

Considering the four-photon situation, the state is
\begin{eqnarray}
(a^\dag_{1H}b^\dag_{1H}+a^\dag_{1V}b^\dag_{1V}+a^\dag_{2H}b^\dag_{2H}+a^\dag_{2V}b^\dag_{2V})^2\vert 0\rangle.
\end{eqnarray}
Here four photons constitute two hyperentangled states entangled both in polarization and spatial modes, which can both be purified with our installation.
However, due to the indistinguishability of photons in one spatial mode, postselection is utilized to select the four-mode coincident cases, where there is one photon in each spatial mode $c_1,c_2,d_1,d_2$. Only in this case can the photons and state be differentiated and used in quantum information process.

Let discuss the purification process considering different error cases. From the analysis in Sec.II we know that the phase-flip errors will be erased in postselection and have no effect on the spatial modes, we only discuss the bit-flip errors in the follows. The effect of PBS can be described as
\begin{eqnarray}
a^\dag_{1H}\rightarrow c^\dag_{1H}, b^\dag_{1H}\rightarrow d^\dag_{1H},\nonumber\\
a^\dag_{1V}\rightarrow c^\dag_{2V}, b^\dag_{1V}\rightarrow d^\dag_{2V}.
\end{eqnarray}
Firstly, in the absence of errors, four-mode coincidence projects the state onto
\begin{eqnarray}
\vert \Psi \rangle_1=(c^\dag_{1H}d^\dag_{1H}+c^\dag_{1V}d^\dag_{1V})(c^\dag_{2H}d^\dag_{2H}+c^\dag_{2V}d^\dag_{2V})\vert 0\rangle,
\end{eqnarray}
a state of two independent polarization-entangled pairs, one in mode $c_1d_1$ and another one in mode $c_2d_2$. And the polarization states are $\vert \Phi^+ \rangle$ as we want.
Suppose a single bit-flip error happens with probability $e$, the case that one of the two pairs catches a bit-flip error occurs with probability $2e(1-e)$. In this case, the state results in three-mode coincident cases and should be thrown away for the aim of purification.
The situation that each pair gets a bit-flip error occurs with probability $e^2$. And then the four-mode coincident case corresponds to the follow state
\begin{eqnarray}
\vert \Psi \rangle_2=(c^\dag_{1H}d^\dag_{2H}+c^\dag_{1V}d^\dag_{2V})(c^\dag_{2H}d^\dag_{1H}+c^\dag_{2V}d^\dag_{1V})\vert 0\rangle,
\end{eqnarray}
which represents two independent polarization entangled state $\vert \Phi^+ \rangle$ in mode $c_1d_2$ and $c_2d_1$.
$\vert \Psi \rangle_1$ and  $\vert \Psi \rangle_2$ can not be discriminated by spatial modes. Therefore, two entangled pairs we get from spatial modes $c_1d_1$ and $c_2d_2$ are mixed ones and the fidelity of $\vert \Phi^+ \rangle$ is
\begin{eqnarray}
F=\frac{(1-e)^2+e^2/4}{(1-e)^2+e^2}.
\end{eqnarray}
Here the term $e^2/4$ in the numerator comes from the case that state $\vert \Psi \rangle_2$ has 1/4 probability be projected onto $\vert \Psi \rangle_1$ by entanglement swapping \cite{swap}. The Ref. \cite{5} did not discuss the two-error case. If the error rate $e$ is low, $e^2$ can be ignored and the fidelity will approach 1 approximatively. And if the error rate is considerable, our protocol will also increase the fidelity of $\vert \Phi^+ \rangle$.
From another perspective we find the mixed state of $\vert \Psi \rangle_1$ and  $\vert \Psi \rangle_2$ is also useful in quantum communication. Based on the principle of entanglement swapping, the Bell state measurement results in $c_1c_2$ and $d_1d_2$ have accurate correspondence, which can be used to generate quantum keys.

Now we discuss the single-pair and two-pair cases together. The two-pair case with no errors and two bit-flip errors may also lead to two-mode coincidences, which are inevitably contribute to the accepted postselection coincidence corresponding to the single-pair case. To rule out this situation, photon number resolving detectors \cite{pnrd,pnrd2} are required to get rid of the cases with more then one photon in one spatial mode. Thus the two-photon cases kept correspond to a pure entangled state.

However, the quality of the entanglement purified by our DEPP with PDC source would be influenced by the photon loss, which is another important error source. Suppose the PDC source creates one photon pair with probability $p$, then the probability of two-pair case is $p^2$. The single photon loss rate is denoted with $m$. In the first place, all the results contribute to a pure maximally entangled state $\vert \Phi^+ \rangle$ for single-pair case in the absences of photon loss, whose probability is $p(1-m)^2$. In the second place, two-pair case with two photon missing may offer two-photon two-mode coincidence which can not be picked out. The probability that two of four photons missing is $6m^2(1-m)^2$, among which the cases two perfectly entangled photons kept, both two photons in maximally mixed state kept and two photons kept in one person's hand occur with equal odds. Other cases with photon number unequal to 2 would be rejected by postselection and photon number measurements. The fidelity of the final state in the case of each party receiving one and only one photon is
\begin{eqnarray}
F'=\frac{p(1-m)^2+ 2p^2m^2(1-m)^2}{p(1-m)^2+ 4p^2 m^2(1-m)^2}
\end{eqnarray}
The fidelity is independent of error rate $e$, which means our DEPP can work in despite of errors. The decline in fidelity is caused by the two-pair case and photon loss. Therefore there is a trade-off between the generation probability $p$ and the fidelity of final states. In order to suppress the impact of the two-pair case, the generation probability $p$ should be restricted, which is usually set to $p \leq 0.1$  \cite{20}. Moreover, $p$ can be adjusted to get the desired fidelity with a given loss rate $m$.

In a word, considering the PDC source and practical transmission with photon loss, our protocol can only get relatively high fidelity entangled states in a probabilistic sense. More purification steps are required to get a nearly perfect pure entangled state. However, the photon loss may also has an impact on other entanglement purification schemes when use currently available PDC source.

\section{discussion and summary}
The task of EPPs is to correct the bit-flip and phase-flip errors in the polarization states.
In conventional EPPs, two less-perfectly entangled pairs are used to detect a bit-flip error by parity check in each step. The case that both pair have bit-flip errors can not be selected and kept for the next round. Hence the errors can not be eliminated completely and the remaining state is also a mixed one. Moreover, the phase-flip errors should be transformed into bit-flip ones and be purified with the same method. With iterative purification steps, the purity and entanglement of the quantum state can be increased. However, the way using two fault pairs to detect errors can never get an indeed pure state.

In Simon's protocol \cite{5}, an additional spatial entanglement is used as quantum resource, which remains perfect pure maximally entangled states during the transmission.
The bit-flip errors can be corrected entirely sacrificing the spatial entanglement. However, the spatial resource is consumed in the first step and then the phase-flip errors can only be corrected in the next step with conventional method. Following this principle, Sheng's protocol utilizes two additional entanglements both in spatial and frequency DOFs, which are used to correct the bit-flip and phase-flip errors one by one. In other words, each error is corrected at the expense of one maximally entanglement.

 In our DEPP, only one additional entanglement is required to correct these two errors. These two kinds of error are transferred to the spatial DOF in one step. The bit-flip errors are corrected and the phase-flip one will be erased in measurements.
Compared with the previous EPPs\cite{1,2,3,4,5,6,7} and DEPP \cite{17}, our scheme has some interesting features as follows: (1) There is no requirement of fidelity on the input states. The proportion of each Bell state in the mixed state represents the probability two parties obtain the desired state in corresponding spatial modes. In other words, our DEPP can always work no matter the noise is. (2) The present scheme can obtain an absolute pure entangled state deterministically. And bit-flip and phase-flip errors can be corrected and eliminated in one step. (3) Only one additional entanglement is required. And the set-up is composed of passive linear optics, which is easy to implement with current techniques.

In summary, we have present a deterministic entanglement purification scheme for arbitrary mixed polarization state with present technology. Two remote parties can obtain a pure maximally entangled polarization state deterministically with the help of spatial entanglement. We also discussed the purification with PDC source and practical transmission, in which relatively high fidelity entangled states can be obtained.
As only passive linear optics is required and the input entangled states is available, this scheme might have good application in quantum repeater and quantum information based on entanglements.

\section*{Acknowledgement}

We thank Dr. X. J. Duan for comments and discussions. This work is supported by the National Natural Science
Foundation of China under Grant No. 11004258, and Fundamental Research Funds for the
Central Universities Project under Grant No. CDJZR10100018.


\begin{thebibliography}{99}

\bibitem{1} C. H. Bennett, G. Brassard, S. Popescu, B. Schumacher, J. A. Smolin, and W. K. Wootters, Phys. Rev. Lett. 76, 722 (1996).



\bibitem{swap} M. Zukowski, A. Zeilinger, M. A. Horne, A. K. Ekert, Phys. Rev. Lett. 71, 4287 (1993).

\bibitem{repe} H. J. Briegel, W. D\"{u}r, J. I. Cirac, and P. Zoller, Phys. Rev. Lett. 81, 5932 (1998).

\bibitem{2} D. Deutsch, A. Ekert, R. Jozsa, C. Macchiavello, S. Popescu and A. Sanpera, Phys. Rev. Lett. 77, 2818 (1996).

\bibitem{3} J. W. Pan, C. Simon, and A. Zellinger, Nature (London) 410, 1067 (2001)

\bibitem{4} J. W. Pan, S. Gasparonl, R. Ursin, G. Weihs, and A. Zellinger, Nature (London) 423, 417 (2003).

\bibitem{5} C. Simon and J. W. Pan, Phys. Rev. Lett. 89, 257901 (2002).

\bibitem{6} Y. B. Sheng, F. G. Deng, and H. Y. Zhou, Phys. Rev. A 77, 042308 (2008).

\bibitem{7} L. Xiao, C. Wang, W. Zhang, Y.D. Huang, J.D. Peng,  G.L. Long, Phys. Rev. A  77, 042315 (2008).

\bibitem{8} C. Wang, Y. B. Sheng, X. H. Li et al., Sci China Ser. E-Tech. Sci., 52(12): 3464 (2009)

\bibitem{17} Y. B. Sheng, F. G. Deng, Phys. Rev. A 81, 032307 (2010).

\bibitem{18} J. L. O'Brien, A. Furusawa, J. vu\u{c}kovic, Nature Photonics, 3, 687 (2009)

\bibitem{bbm92} C. H. Bennett, G. Brassard, N. D. Mermin, Phys. Rev. Lett 68, 557 (1992).

\bibitem{pnrd} E. Pomarico, B. Sanguinetti, R. Thew, H. Zbinden, Opt. Expre., 18, 10750-10759 (2010).

\bibitem{pnrd2} X. L. Chen, E. Wu, L. L. Xu, Y. Liang, G. Wu, H. P. Zeng, Appl. Phys. Lett., 95, 131118 (2009).

\bibitem{effect} A. K. Jha, G. A. Tyler, R. W. Boyd, Phys. Rev. A 81, 053832 (2010).

\bibitem{phase} J. Min\'{a}\u{r}, H. deRiedmatten, C. Simon, H. Zbinden, N. Gisin, Phys. Rev. A 77, 052325 (2008).

\bibitem{20} J. Fulconis, O. Alibart, J. L. O'Brien, W. J. Wadsworth, J. G. Rarity, Phys. Rev. Lett 99, 120501 (2007).




\end{thebibliography}
\end{document}